\documentclass[prd,showpacs,superscriptaddress,nofootinbib,floatfix,twocolumn
]{revtex4}
\usepackage{graphicx,multirow,color,xspace}
\usepackage{amsfonts,amsmath,amssymb}
\usepackage[innercaption]{sidecap}
\usepackage{epstopdf}
\usepackage{hyperref}
\graphicspath{{figures/}}
\begin{document}
\setcounter{topnumber}{1}
\title{Hawking radiation and the boomerang behaviour of massive modes near a horizon}
\author{G. Jannes}
\affiliation{Universit\'{e} de Nice Sophia Antipolis, Laboratoire J.-A. Dieudonn\'{e}, UMR
CNRS-UNS 6621, Parc Valrose, 06108 Nice Cedex 02, France}
\affiliation{Low Temperature Laboratory, Aalto University School of Science, PO Box 15100, 00076 Aalto, Finland}
\author{P. Ma\"issa}
\affiliation{Universit\'{e} de Nice Sophia Antipolis, Laboratoire J.-A. Dieudonn\'{e}, UMR
CNRS-UNS 6621, Parc Valrose, 06108 Nice Cedex 02, France}
\author{T. G. Philbin}
\affiliation{School of Physics and Astronomy, University of St Andrews, North Haugh, St
Andrews, Fife KY16 9SS, Scotland, UK}
\author{G. Rousseaux}\email{Germain.Rousseaux@unice.fr}
\affiliation{Universit\'{e} de Nice Sophia Antipolis, Laboratoire J.-A. Dieudonn\'{e}, UMR
CNRS-UNS 6621, Parc Valrose, 06108 Nice Cedex 02, France}
\date{\today}

\begin{abstract}
We discuss the behaviour of massive modes near a horizon based on a study of the dispersion relation and wave packet simulations of the Klein-Gordon equation. We point out an apparent paradox between two (in principle equivalent) pictures of black hole evaporation through Hawking radiation. In the picture in which the evaporation is due to the emission of positive-energy modes, one immediately obtains a threshold for the emission of massive particles. In the picture in which the evaporation is due to the absorption of negative-energy modes, such a threshold apparently does not exist. We resolve this paradox by tracing the evolution of the positive-energy massive modes with an energy below the threshold. These are seen to be emitted and move away from the black hole horizon, but they bounce back at a ``red horizon'' and are re-absorbed by the black hole, thus compensating exactly for the difference between the two pictures. For astrophysical black holes, the consequences are curious but do not affect the terrestrial constraints on observing Hawking radiation. For analogue gravity systems with massive modes, however, the consequences are crucial and rather surprising.
\end{abstract}
\maketitle

\section{Introduction}
\subsection{Black hole evaporation through the emission of positive-energy Hawking modes}
The evaporation of black holes through Hawking radiation is one of the cornerstones of post-classical gravity\cite{Hawking:1974sw,Birrell:1982ix,Fabbri:2005mw}. In Hawking's original derivation~\cite{Hawking:1974sw} for the case of a gravitational collapse, the emphasis lay on the positive-energy modes that cross the horizon just before it is actually formed, escape from the black-hole spacetime and can in principle be detected by an asymptotic observer. Hawking already observes that the decrease of the black hole mass, and the accompanying decrease of the area of the event horizon ``must, presumably, be caused by a flux of negative energy
across the event horizon which balances the positive energy flux emitted to
infinity. One might picture this negative energy flux in the following way. Just
outside the event horizon there will be virtual pairs of particles, one with negative
energy and one with positive energy. The negative particle \dots can tunnel through the event horizon to the region inside the black hole \dots In this region the particle can exist as a real particle with a timelike momentum vector even though its energy relative to infinity as measured by the
time translation Killing vector is negative. The other particle of the pair, having
a positive energy, can escape to infinity where it constitutes a part of the thermal
emission described above.'' Hawking warns, however, that it ``should be emphasized that these pictures of the mechanism responsible for the thermal emission and area decrease are heuristic only
and should not be taken too literally.'' 

With regard to massive particles, Hawking notes that 
``As [the black holes] got smaller, they would get hotter and so would radiate faster. As the temperature rose, it would exceed the rest mass of particles such as the electron and the muon and the black hole would begin to emit them also.'' Therefore, ``the rate of particle
emission in the asymptotic future \dots will again be that of a body with temperature $\kappa/2\pi$. The only difference from the zero rest mass case is that the frequency $\omega$ in the thermal factor $(\exp(2\pi\omega\kappa^{-1})\mp 1)^{-1}$ now includes the rest mass
energy of the particle. Thus there will not be much emission of particles of rest
mass $m$ unless the temperature $\kappa/2\pi$ is greater than $m$.'' A similar conclusion was reached in~\cite{Page:1976df}. Indeed, for the black hole to emit for example an electron\footnote{We neglect complications~\cite{Page:1977um} due to the electrical charge.}, its temperature must be on the order of $T= 10^9$ Kelvin.\footnote{From $mc^2=k_B T$ and using $m_e= 9.11\times 10^{-31} \mathrm{kg}$, one obtains $T=5.93\times 10^9$ Kelvin. This rough estimate is in agreement with \cite{Birrell:1982ix} (see p.261). The associated frequency, from $\hbar\omega = mc^2$, is $\omega _c = 7.76\times 10^{20} \mathrm{Hz}$.} 
Due to emission of massless particles the black hole will eventually become small enough for the temperature to reach $10^9$ K, and then the radiation will contain electrons and positrons. But for most of the lifetime of the black hole, the mass cutoff prevents any (significant) radiation of electrons.

Implicit in the above reasoning is that the black hole emission corresponds (by definition) to what can be detected by an asymptotic observer. This definition makes sense from a relativistic point of view because a black hole in relativity is, strictly speaking, not just the central object enclosed by a horizon, but also the entire spacetime surrounding it. This definition however has various well-known complications, for example that, strictly speaking, one needs an infinite amount of time to ascertain the existence of a black hole. In an astrophysical sense, therefore, the alternative definition in which the black hole is only what lies beyond the horizon makes much more sense. In the context of analogue gravity, which we will discuss in section~\ref{S:analogue-gravity}, 
only the horizons are the essential ingredient and it makes little sense to speak of asymptotic observers. One can already anticipate that, from this second point of view, what is emitted by the black hole (i.e., what moves away from the horizon) is not necessarily equal to what is detected at infinity. As we will see, the difference can actually be substantial.

\subsection{Black hole evaporation through the absorption of negative-energy Hawking modes}
Hawking's heuristic picture in which the black hole shrinks through the absorption of negative-energy modes was given a physical basis in the following context. An important problem with Hawking's calculation is the so-called ``transplanckian problem'': the exponential frequency-shift near the horizon means that particles detected just seconds after the formation of the black hole seemingly originate from modes with an energy exceeding by far the total energy of the observable universe just before the formation of the black hole. In 1981, in what is usually considered the founding paper of analogue gravity (see~\cite{Barcelo:2005fc} for a review), Unruh~\cite{Unruh:1980cg} noted that the propagation of sound waves in a fluid with a subsonic-to-supersonic transition is formally identical to the propagation of scalar fields in a black hole spacetime. Since the continuum description of fluid systems has a natural cut-off (the intermolecular distance), the transplanckian problem of Hawking radiation can be studied in this fluid-analogy context. Unruh took up his own suggestion and studied the (typically ``subluminal'') modification of the dispersion relation for sound waves  compared to a ``relativistic'' or acoustic one, and its influence on the Hawking spectrum~\cite{Unruh:1994je}. For such subluminal dispersion, an outgoing positive-energy Hawking packet can be traced back to a mixed positive/negative-energy mode originating from \emph{outside} the horizon. If the dispersion is changed to superluminal then the positive-energy mode originates from \emph{inside} the horizon, but this difference does not alter the size of the Hawking effect compared to the subluminal case. This suggests that the evaporation (loss of mass) of the black hole does not rely so much on this positive-energy emission, but instead has its physical origin in the absorption of the negative-energy partner. The negative-energy mode is always consumed by the black hole, regardless of the details of the high-energy dispersion.

It is far beyond our intention to clear up the exact physical origin of Hawking emission and black-hole evaporation. What we do want to point out, however, is the following. The emission of massive positive-energy modes immediately seems to be limited by a lower cut-off, as pointed out above, since the energy of a massive mode in the flat space-time of an asymptotic observer cannot be less than its mass-energy. This is not the case, however, for the absorption of massive negative-energy modes, since these fall into the black hole and so never obey the flat space-time dispersion relation. In fact, as we will discuss in detail, the process of creation of positive/negative-energy pairs near the horizon is not restricted by the mass of the corresponding modes. Massive negative-energy modes are created and absorbed by the black hole according to exactly the same mechanism and at exactly the same rate\footnote{Again, modulo complications~\cite{Page:1977um} due to, e.g., the electrical charge. We do not consider the Dirac equation, although the dispersion relation discussed below also applies in that case.} as massless modes. Since the two pictures of black hole evaporation through Hawking radiation (either through the emission of positive-energy modes or through the absorption of negative-energy modes) should in principle lead to exactly the same evaporation rate, this raises an apparent paradox. The paradox is qualitatively easy to solve: all positive-energy modes emitted with energy (relative to asymptotic infinity) below the rest-mass energy must somehow be reabsorbed by the black hole in order to counterbalance the ``excess'' negative-energy absorption. A detailed analysis of how this process takes place leads to some curious consequences, and in particular to several potentially crucial aspects for experiments in analogue gravity systems with massive modes.

Before starting our analysis, note that we will talk about massive \emph{modes} to avoid any caveats concerning the definition of \emph{particles} in a curved spacetime. This is not just a matter of terminology. Strictly speaking, particles do not go on-shell until they have reached the flat region of spacetime (either at infinity, or when they are trapped by the potential well of a particle detector before reaching infinity). Care should therefore be taken when reasoning in terms of particles. For example, the usual heuristic argument about particle pair creation near the horizon (due to the effect of curvature on a distance of the order of the Compton wavelength) might induce one to think that massive particle pairs are created near a black hole horizon only when the action is sufficient to provide an energy at least equal to the pair's mass. This argument might be useful in that it correctly reproduces the threshold for particles to appear at infinity, as mentioned above. However, in order to make estimates about what happens anywhere between the black hole horizon and infinity, it can be quite deceptive, and it is more indicated to reason in terms of modes or wave packets, as we will do. For analogue gravity systems, such complications do not usually arise because the emitted modes are in general phonons or other collective excitations, which are in any case naturally decribed in terms of waves rather than particles. Moreover, the configurations considered for experimental detection of analogue Hawking radiation are typically stationary and have an asymptotically constant background flow velocity $U$. The ambiguity of the quantum vacuum definition in a non-stationary curved spacetime~\cite{Birrell:1982ix} does therefore not arise here (see~\cite{Macher:2009nz} for useful remarks in the case of phonons in a BEC). We will come back extensively to the question of detectability in sections~\ref{S:HR} and \ref{S:analogue-gravity}.

\section{Dispersion relation for massive modes in a black hole space-time}
The general dispersion relation for massive Klein-Gordon modes trying to escape a black hole can be written
\begin{equation}\label{disp-rel-squared}
(\omega-Uk)^2=m^2+c^2k^2
\end{equation}
Note that, strictly speaking, the mass $M$ of the mode is given by $M=m\hbar/c^2$. We set $\hbar=1$ and will refer throughout to $m$ as the mass. As usual, $U<0$ can either represent the velocity of an observer freely falling into the black hole (in the case of gravity), or the velocity of a background flow against which the mode counter-propagates (for analogue gravity systems). Both cases are described by exactly the same Painlev\'e-Gullstrand-Lema\^itre metric
\begin{equation} \label{metric}
ds^2=[c^2-U(x)^2]dt^2 - 2U(x) dt\, dx -dx^2
\end{equation}
where $\hat{x}$ is the direction perpendicular to the horizon (which contains all the essential features that we are interested in\footnote{ Note that a bouncing effect similar to what we will describe also exists for (even massless) modes with non-zero angular momentum, see e.g.~\cite{Jacobson:2007jx}, as is well known. The crucial difference is that the bouncing which we will describe is valid even for modes that move away from the black hole in a purely radial fashion. It is therefore truly a consequence of the mass of the modes.}), and $c$ represents the speed of light, or in analogue gravity the (low-frequency) speed of sound or any other characteristic velocity of the system which leads to an effective relativistic behaviour (typically) at low frequencies. The strict equivalence (at least at low frequencies) means that we do not have to worry about terminology, and so we will typically refer quite generally to, for instance, $|U|=c$ as the ``sonic'' case, $U$ as the ``background flow'', and ``counter-propagating modes'' for modes that try to escape the black hole (or enter a white hole), as if they were moving against a background flow.

The action of the Klein-Gordon field $\phi$ in the metric (\ref{metric}) is
\begin{equation}  \label{action}
S=\frac{1}{2}\int dt\,dx\left[\frac{1}{c^2}\left|\partial_t\phi+U\partial_x\phi\right|^2-|\partial_x\phi|^2-\frac{m^2}{c^2}|\phi|^2\right],
\end{equation}
which gives the field equation
\begin{equation}  \label{KGeqn}
\partial_t(\partial_t\phi+U\partial_x\phi)+\partial_x(U\partial_t\phi+U^2\partial_x\phi)-c^2\partial_x^2\phi+m^2\phi=0
\end{equation}
and the dispersion relation (\ref{disp-rel-squared}). Wave packets obeying the Klein-Gordon equation (\ref{KGeqn}) possess two conserved quantities due to the invariance of the action (\ref{action}) under 1)  the transformation $\phi\rightarrow e^{i\alpha}\phi$, $\alpha$ constant, and  2) time translation (for time-independent $U$). The former invariance gives conservation of the Klein-Gordon norm
\begin{equation} \label{norm}
N=\frac{i}{2c^2}\int_{-\infty}^\infty dx\left[\phi^*(\partial_t\phi+U\partial_x\phi)-\phi(\partial_t\phi^*+U\partial_x\phi^*)\right],
\end{equation}
whereas the latter gives conservation of (pseudo-)energy
\begin{equation} \label{E}
E=\frac{1}{2}\int_{-\infty}^\infty\!\! dx\!\left[\frac{1}{c^2}|\partial_t\phi|^2+(1-U^2/c^2)|\partial_x\phi|^2+\frac{m^2}{c^2}|\phi|^2\right].
\end{equation}
For waves packets confined to a region where the flow velocity $U$ is constant, the Klein-Gordon norm (\ref{norm}) can be written in $k$-space in terms of the Fourier transform $\tilde{\phi}(k)$ as
\begin{equation} \label{normk}
N=\frac{1}{c^2}\int_{-\infty}^\infty dk(\omega-Uk)|\tilde{\phi}(k)|^2,
\end{equation}
while in similar circumstances the energy (\ref{E}) takes the form
\begin{equation} \label{Ek}
E=\frac{1}{c^2}\int_{-\infty}^\infty dk\,\omega(\omega-Uk)|\tilde{\phi}(k)|^2.
\end{equation}
The quantity that determines the sign of the norm (\ref{norm}) of a wave is the sign of its frequency $\omega-Uk$ in a frame co-moving with the background flow $U$. For waves with positive values of the frequency $\omega$ in the black-hole/lab frame, the sign of the (pseudo-)energy (\ref{E}) is also given by the sign of the co-moving frequency $\omega-Uk$.

A key point is that the dispersion relation (\ref{disp-rel-squared}) far away from the black hole (in Minkowski space-time) is
\begin{equation}\label{disp-rel-Minkowski}
\omega^2=m^2+c^2k^2,
\end{equation}
which does not allow any particles with $\omega<m$. However, sufficiently close to the black hole horizon $U\to -c$, so from (\ref{disp-rel-squared}) massive modes can in principle be created even at frequencies $0 < \omega < m$. 

\begin{figure}
\includegraphics[width=.45\textwidth]{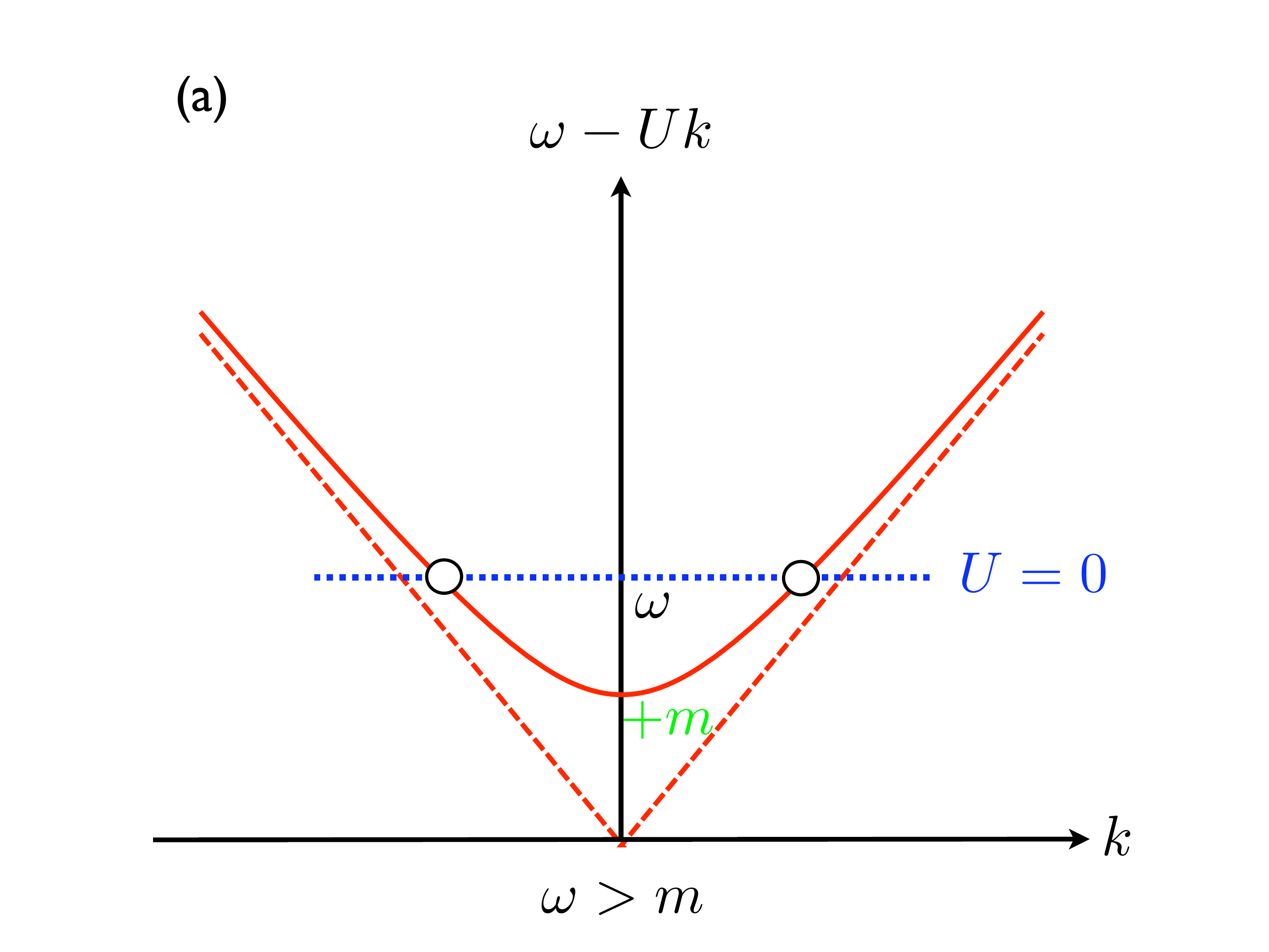}
\includegraphics[width=.45\textwidth]{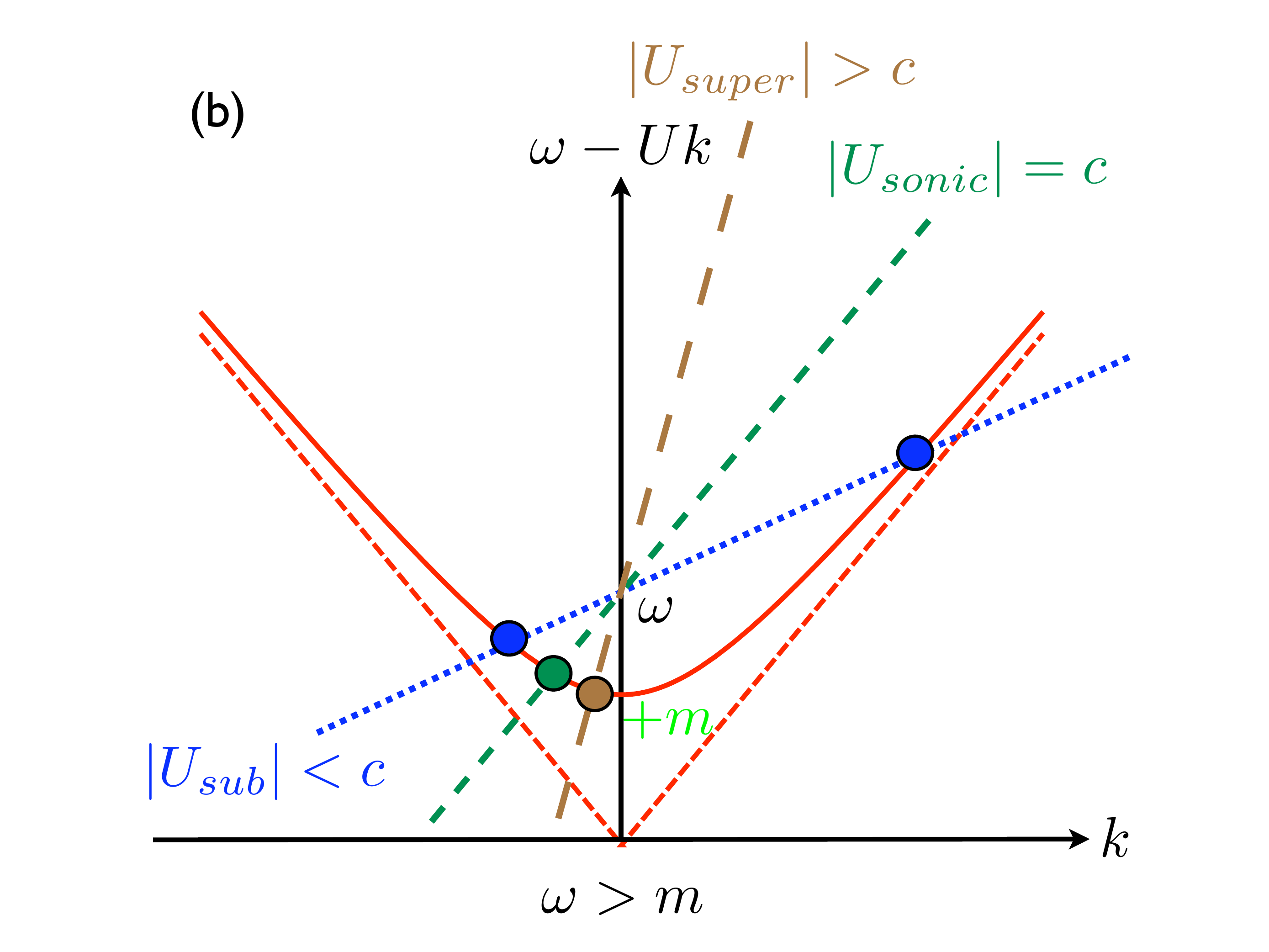}
\caption{\label{Fig:dispersionKGU=0} 
Positive-frequency part of the dispersion relation~\eqref{massive-disp-rel} 
for $\omega>m$ and (a) $U=0$, (b) various values of $U < 0$. $k<0$ corresponds to co-propagating modes, $k>0$ to counter-propagating ones. The massless sonic branches $\omega-Uk=\pm ck$ are indicated (dashed red lines) for comparison. Qualitatively, the $\omega>m$ massive case is identical to the massless case.
}
\end{figure}

\begin{figure} 
\includegraphics[width=.45\textwidth]{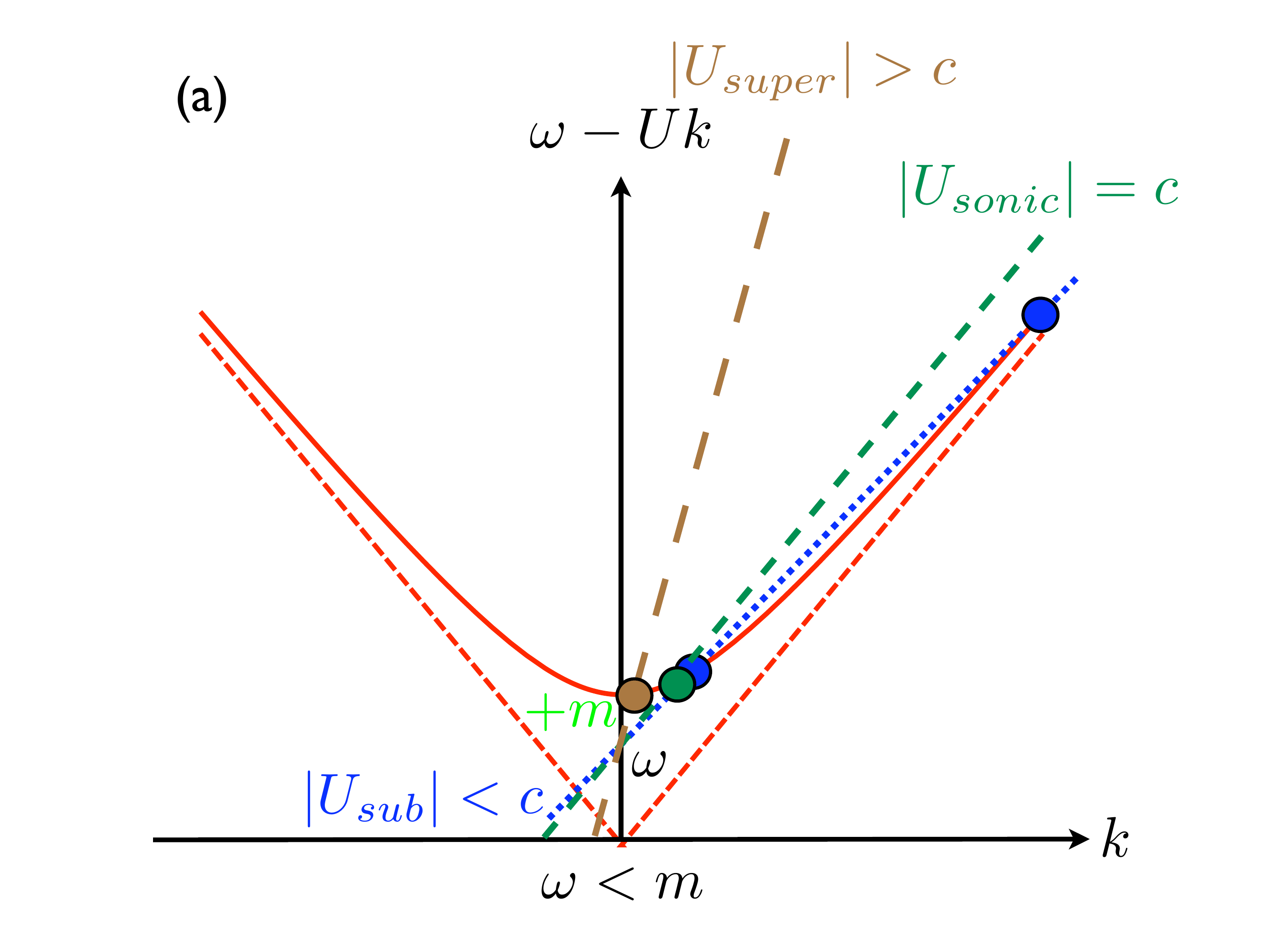}
\includegraphics[width=.45\textwidth]{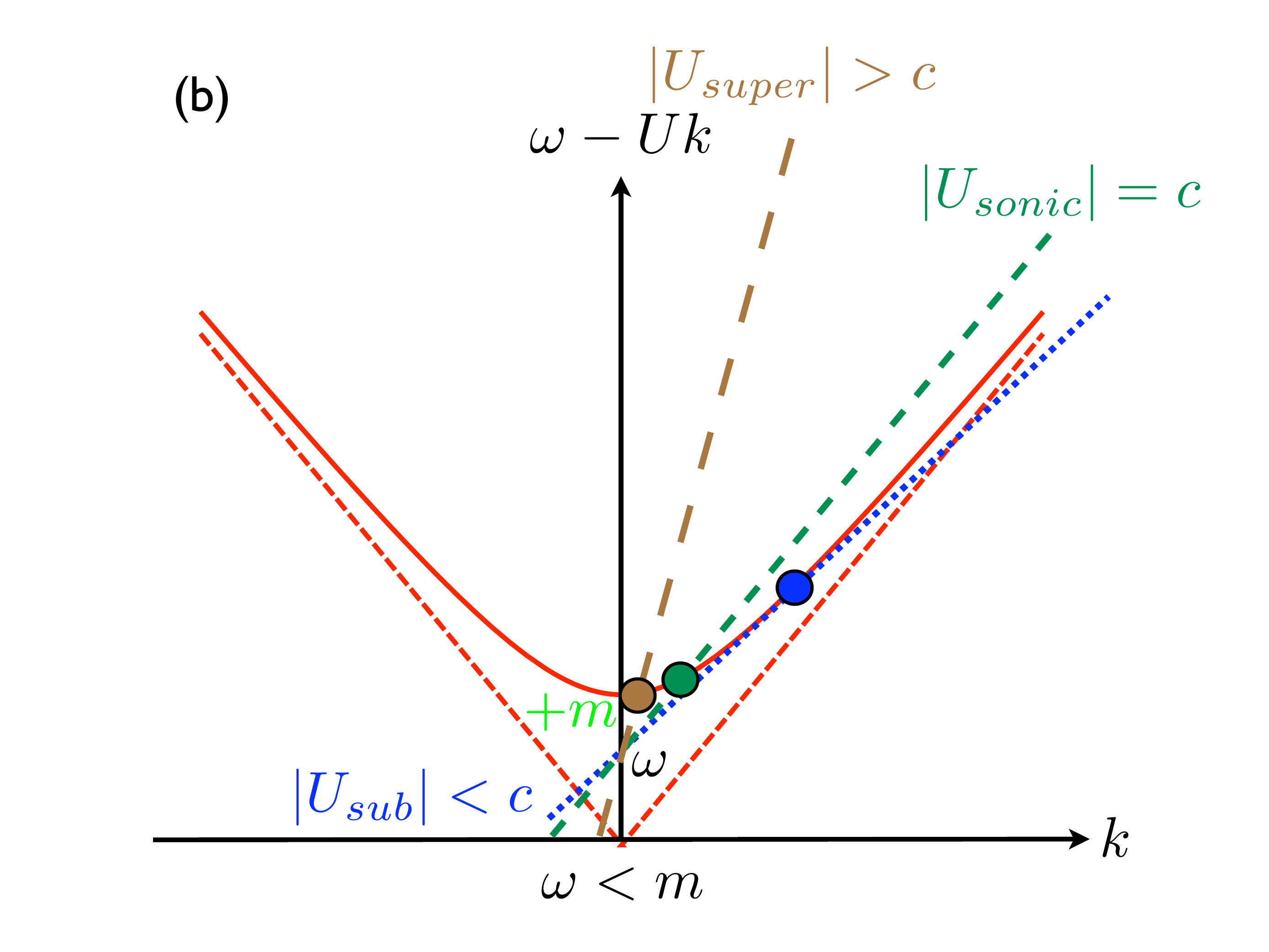}
\caption{\label{Fig:dispersionKGUneq0<} 
Positive-frequency part of the dispersion relation~\eqref{massive-disp-rel} 
for $0<\omega<m$ and various values of $U < 0$. For the value of $|U_\text{sub}|<c$ shown in (a), there are two counter-propagating ($k>0$) modes, one with a positive group velocity $d\omega/dk>0$ in the black-hole/lab frame, the other with a negative group velocity in this frame. In (b), $U_\text{sub}=U^*$, the critical value at which these two solutions merge and only one (double) root $k^*$ remains. For $|U_\text{sub}|<|U^*|$, there are no longer any positive-frequency modes. $U^*$ is therefore characteristic of a turning point or ``red horizon'' (see main text). The massless sonic branches $\omega-Uk=\pm ck$ are again indicated (dashed red lines) for comparison. Note that all co-propagating modes ($k<0$) have disappeared, while the number of counter-propagating modes ($k>0$) depends on the value of $U$.}
\end{figure}

\begin{figure} 
\includegraphics[width=.45\textwidth]{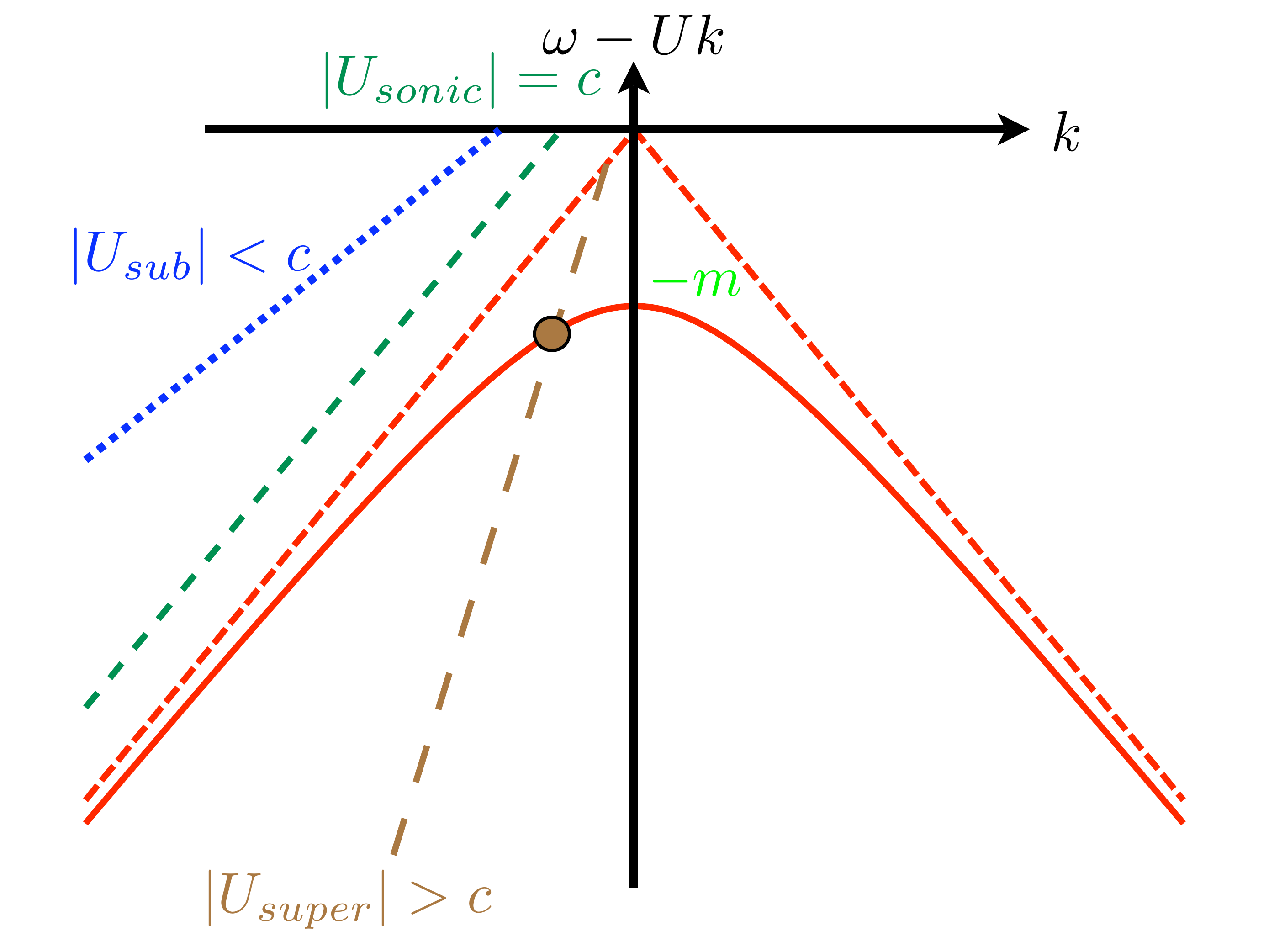}
\caption{\label{Fig:dpn} 
Negative-frequency part of the dispersion relation~\eqref{massive-disp-rel} for $\omega>0$. The behaviour is independent of whether $\omega<m$ or $\omega>m$, and is qualitatively identical to the massless case (in dashed red line). For $|U_\text{sub}|<c$ and $|U_\text{sonic}|=c$, there are no negative-frequency solutions. A single counter-propagating negative-frequency root appears for $|U_\text{super}|>c$.
}
\end{figure}

We will now plot the dispersion relation~(\ref{disp-rel-squared}), i.e.
\begin{equation}\label{massive-disp-rel}
 \omega-Uk=\pm\sqrt{m^2 + c^2k^2},
\end{equation}
for values of $\omega>0$ greater than or less than the mass $m$. For fixed $\omega$ the left- and right-hand sides of (\ref{massive-disp-rel}) are plotted as functions of $k$; the intersection points of these plots are the solutions of the dispersion relation. Figures~\ref{Fig:dispersionKGU=0} and \ref{Fig:dispersionKGUneq0<} represent the positive-frequency branch of the dispersion relation (plus sign in (\ref{massive-disp-rel})), while Fig.~\ref{Fig:dpn} depicts the negative-frequency branch (minus sign in (\ref{massive-disp-rel})). Note that ``positive/negative-frequency'' always refers to the \emph{comoving} frequency ($\omega-Uk$), and therefore coincides with the concept of positive/negative norm, as just mentioned. We always draw $U<0$ (a flow moving to the left) and will focus on the modes counter propagating against the flow, i.e. modes moving to the right in a frame co-moving with the flow. The counter-propagating modes have positive phase and group velocities in the frame comoving with the flow: $(\omega-Uk)/k>0$, $d(\omega-Uk)/d k>0$. Our interest is in counter-propagating modes because these modes try to escape the black hole (or enter a white hole).

\subsection{$\omega>m$ (positive-frequency part)}
The case $\omega>m$ is very similar to the standard massless case, see Fig.~\ref{Fig:dispersionKGU=0}. In the absence of a counter-flow, i.e. far away from the black hole (Fig.~\ref{Fig:dispersionKGU=0}(a)), there are two positive-frequency roots $k_1$ and $k_2=-k_1$ with the same modulus and opposite direction of propagation. When $U\neq 0$ (Fig.~\ref{Fig:dispersionKGU=0}(b)), the situation depends on the value of $U$: for $|U_\text{sub}|<c$, there are two positive-frequency roots $k_1<0$ and $k_2\neq -k_1$; for $|U_\text{sonic}|=c$ and $|U_\text{super}|>c$, only one positive-frequency root remains, namely the  mode $k_1<0$ co-propagating with the background flow $U$. For $|U_\text{super}|>c$, the second root $k_2$ has actually been converted into a counter-propagating negative-frequency root, see Fig.~\ref{Fig:dpn}. This conversion from positive to negative frequency across the horizon is precisely the essence of the Hawking mechanism (at least in the case where there are no higher-order $k$-terms in the dispersion relation). For $|U_\text{sonic}|=c$, $k_2$ has ``disappeared'': $k_2 \to +\infty$ when $|U|\to c^-$, and reappears as $k_2 \to -\infty$ when $|U|\to c^+$. This infinite frequency-shift at the horizon ($|U_\text{sonic}|=c$) is characteristic of the transplanckian problem, which (just like in the massless relativistic case) is present in our problem since we do not consider higher-order $k$-terms in the dispersion relation in our present discussion.
Note that the exact values of the different roots will be slightly different compared to the massless case, but in qualitative terms the behaviour for the positive-frequency modes associated with $\omega>m$ is identical to the massless case. 

\subsection{$\omega<m$ (positive-frequency part)}
For $\omega<m$, important differences with the massless case show up, see Fig.~\ref{Fig:dispersionKGUneq0<}. For $U=0$ (not shown), there are no positive-frequency $\omega<m$ solutions to the dispersion relation, in accordance with~(\ref{disp-rel-Minkowski}).  For $|U_\text{sonic}|=c$ and $|U_\text{super}|>c$, there is only one positive-frequency root. Curiously, and contrarily to the case $\omega>m$, the single remaining mode here is a counter-propagating mode, while the mode co-propagating with the background flow is forbidden. For $|U_\text{sub}|<c$, the number of roots depends on the value of $U$: for sufficiently large $|U|$ (Fig.~\ref{Fig:dispersionKGUneq0<}(a)), there are two positive-frequency roots. These are both counter-propagating, but only one of them (with the higher $k$) has a positive group velocity $d\omega/dk>0$ in the black-hole/lab frame, while the other one is dragged along by the background flow and has negative group velocity relative to the black hole. For a critical value  $U^*$ of $U$ (Fig.~\ref{Fig:dispersionKGUneq0<}(b)), these two solutions merge and only one (double) root $k^*$ remains. For even smaller values of $|U|$, as in the case of $U=0$, there are no longer any positive-frequency solutions. These two $k>0$ roots that merge at $k^*$ correspond to the same counter-propagating mode; it propagates away from the horizon (positive group velocity), experiences a turning point when the flow has decreased to $U^*$, and is dragged back into the black hole (negative group velocity). The positive-frequency mode with $0<\omega<m$ thus exhibits a boomerang trajectory: after being emitted away from the horizon it returns, and falls into the black hole. This boomerang behaviour is further illustrated below by a ray plot and a wave packet simulation.

\subsection{Negative-frequency part (all values of $\omega$)}
The negative-frequency part is shown in Fig.~\ref{Fig:dpn}. The behaviour of the negative-frequency modes is independent of the value of $\omega>0$. It is essentially the same as in the massless case: inside the horizon (where $|U|=|U_\text{super}|>c$), a single negative-frequency solution exists. Outside the horizon ($|U|\leq c$), there are no negative-frequency solutions. This corresponds precisely to the ``disappearance'' of the positive-frequency $k_2$-solution in the massless and $\omega>m$ cases: $k_2$ converts from a positive-frequency root into a negative-frequency root across the horizon.

The key point, however, is that this same appearance of a negative-frequency root when $|U|>c$ is also valid for the case $\omega<m$. Moreover, since this negative-frequency mode is absorbed by the black hole and hence moves into a region where $|U|$ increases, its existence is allowed at all times during its subsequent evolution.

\section{Determination of $U^*$}
We have seen graphically that in the case $\omega<m$, there exists a critical background flow velocity $U^*$ such that, for $|U|>|U^*|$, there are two (counter-propagating) positive-frequency solutions, whereas for $|U|<|U^*|$, both these solutions disappear. For $|U|=|U^*|$, there is therefore a double root $k^*$ or a saddle-node bifurcation in dynamical-systems language~\cite{Nardin,Rousseaux:2010md}. 

For $U^2\neq c^2$, the dispersion relation (\ref{disp-rel-squared}) can be written as 
 \begin{equation}
  k^2-\frac{2\omega U}{U^2-c^2}k+\frac{\omega^2-m^2}{U^2-c^2}=0,
 \end{equation}
with a discriminant 
 \begin{equation}
  \Delta=\frac{4\omega^2U^2}{(U^2-c^2)^2}-4\frac{\omega^2-m^2}{(U^2-c^2)^2}(U^2-c^2).
 \end{equation}
There will be real solutions for $k$ as long as $\Delta\geq 0$, i.e. $ U^2 \geq c^2\left[1-\left(\omega/m\right)^2\right]$. The general solution for the roots $k_{1,2}$ is then
\begin{equation}
 k_{1,2}=\frac{\omega U}{U^2-c^2}\left(1\pm \sqrt{1-\frac{(\omega^2-m^2)(U^2-c^2)}{\omega^2U^2}}\right).
\end{equation}
The critical value $U^*$ can be obtained by requiring $\Delta$ to vanish. 
This can only be satisfied when $\omega<m$, and gives 
\begin{equation}\label{U*}
 U^*=\pm c\sqrt{1-\left(\frac{\omega}{m}\right)^2},
 \end{equation}
 with a corresponding critical wavenumber
\begin{equation} \label{k*}
k^* =\mp \frac{m^2}{\omega c}\sqrt{1-\left(\frac{\omega}{m}\right)^2}.
\end{equation}
As we consider flows $U(x)<0$, the lower signs apply in (\ref{U*}) and (\ref{k*}). The critical value (\ref{U*}) occurs outside the black-hole horizon $U=-c$ and corresponds to a turning point for the counter-propagating $\omega<m$ mode as it tries, and fails, to escape the pull of the black hole.

The case $U=\pm c$ gives in the dispersion relation (\ref{disp-rel-squared})
\begin{equation} \label{ksol1}
 k=\pm \frac{\omega^2-m^2}{2\omega c}
\end{equation}
so that only one solution remains (for $U<0$). The sign of the single root depends on the relation between $\omega$ and $m$. Taking $U=U_\text{sonic}=-c$, the solution (\ref{ksol1}) has $k<0$ for $\omega>m$, corresponding to a co-propagating mode crossing the horizon into the black hole. For $\omega<m$, $k>0$ in (\ref{ksol1}), corresponding to a counter-propagating mode being dragged across the horizon after encountering a turning point while trying to escape to infinity. These quantitative results are in agreement with Figs.~\ref{Fig:dispersionKGU=0}(b) and \ref{Fig:dispersionKGUneq0<}.

If $\omega=m$, then 
\begin{equation}
 k\left[k-\frac{2mU}{U^2-c^2}\right]=0
\end{equation}
and the case is qualitatively the same as $\omega>m$, except that there is always one solution at $k=0$.

\begin{figure}   
\includegraphics[width=.45\textwidth]{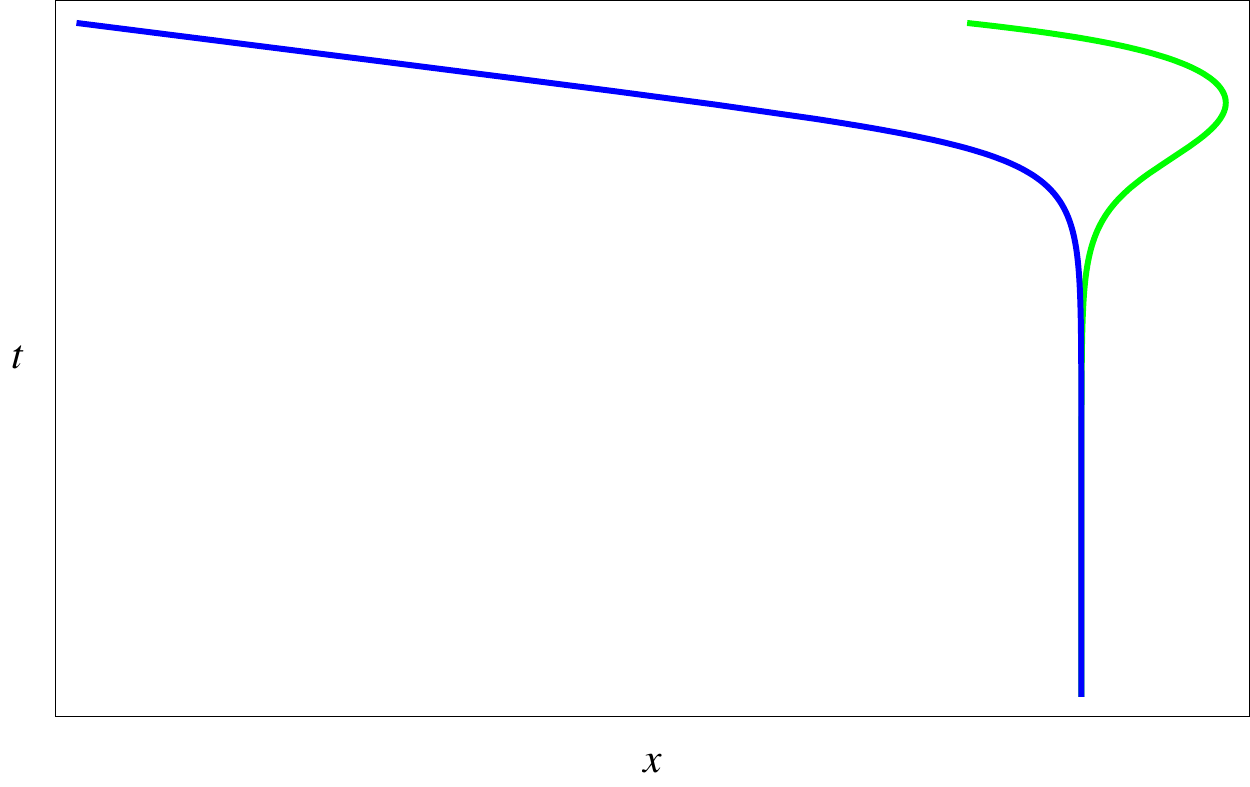}
\includegraphics[width=.45\textwidth]{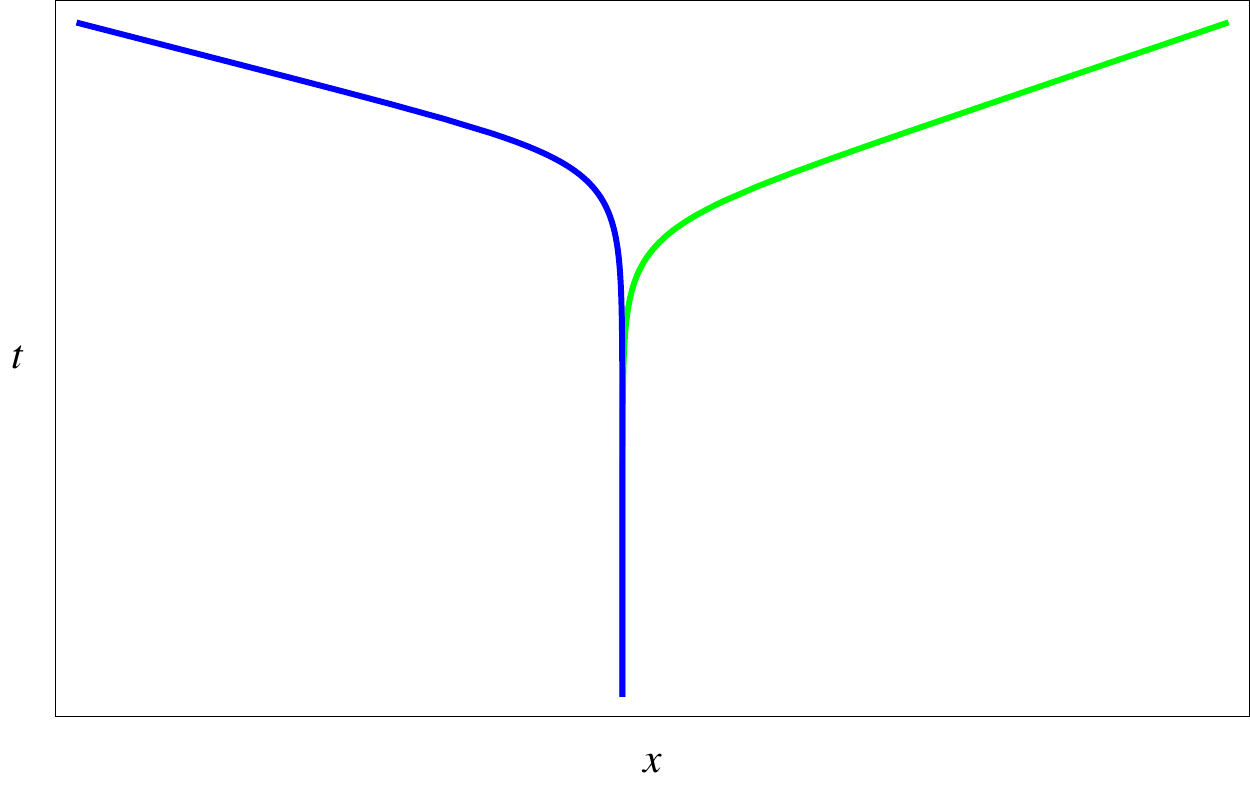}
\caption{\label{Fig:ray-plots} 
Ray trajectories for $0<\omega<m$ (top) and $\omega>m$ (bottom) modes trying to escape from a black hole on both sides of the horizon (green: outside the horizon, blue: inside the horizon). After remaining extremely close to the horizon for some time, the modes that started off just inside the horizon (blue lines) are dragged into the interior of the black hole. The modes just outside the horizon eventually manage to escape at least some distance from the black hole. For the case $\omega>m$ (green ray on the bottom), they escape to infinity. For $\omega<m$ (green ray on the top), the mode escapes some distance from the horizon, but reaches a turning point or ``red horizon'', bounces back and is re-absorbed. The ray trajectories can also be interpreted as corresponding to modes trying to penetrate a white hole, by inversing the time direction. The flow then reverses direction ($U>0$) and the interior of the white hole is on the left of the figures. The counter-propagating rays pile up on the horizon from both sides (green: from the outside, blue: from the inside). Note that, for the case $\omega<m$ (top), the green ray originates inside the white hole, is ejected, turns around at the ``blue horizon'' (the time-reverse of the red horizon, see also main text), tries to enter the white hole again but is unable to penetrate across the white-hole horizon where it piles up. 
}
\end{figure}

\section{Ray plots and wave packet simulation}
We will now plot ray trajectories for the cases $\omega<m$ and $\omega>m$. This will confirm that, for $\omega<m$, there is a (mass- and frequency-dependent) turning point, where the outgoing positive-frequency mode is blocked, bounces back towards the horizon and is reabsorbed by the black hole.

\subsection{Ray trajectories}
We plot  in Fig.~\ref{Fig:ray-plots} ray trajectories for counter-propagating modes that start off infinitesimally close to the horizon on both sides of it. The mode starting off on the inside eventually falls down into the interior of the black hole, independently of its frequency $\omega$, since $|U|>c$ inside the horizon. For modes starting just outside the horizon, the behaviour is completely different depending on whether $\omega>m$ or $\omega<m$. 

For $\omega>m$, as in the massless case, modes starting off just outside the horizon eventually escape to infinity. Note that the group velocity of the massive mode is slightly smaller than that of its massless counterpart, and the difference increases towards low $k$, as can be seen graphically from the slope of $\omega-Uk$ in  Fig.~\ref{Fig:dispersionKGU=0}. However, on the outside of the horizon, $|U|<c$, and so $d\omega/d k > |U|$, as can again easily be confirmed graphically from Fig.~\ref{Fig:dispersionKGU=0}. This can be interpreted as follows: the mode starts off with a very high wavenumber, at which the group velocity $d\omega'/dk$ in a frame comoving with the flow ($\omega'=\omega-Uk$) barely differs from the massless case, i.e. $d\omega'/d k (x=x_H+\epsilon)\approx c>U(x=x_H+\epsilon)$, with $x_H$ the position of the horizon and $\epsilon$ a small positive quantity. The mode therefore gradually escapes, and is redshifted (towards lower $k$) in the process. The decrease in $|U|$ as the ray moves away from the horizon is more important than the decrease in $d\omega'/d k$ (graphically: the slope $d\omega'/d k$ is always larger than $|U|$), and so the massive mode indeed escapes.

For $\omega<m$, on the other hand, as the massive mode $k_1$ moves away from the horizon and is redshifted in the process, it will reach the critical $U^*$ beyond which there are no longer any positive-energy solutions for $k$. The process here is typical for a saddle-node bifurcation: at $U^*$ the two roots $k_1$ (with positive group velocity $c_g=d\omega/dk$) and $k_2$ (with negative $c_g$) merge and $c_g$ vanishes. The outgoing ray is therefore blocked and turns back towards the black hole, in a boomerang fashion. In other words, the outgoing ($c_g>0$) mode $k_1$ is converted (red-shifted) into an ingoing ($c_g<0$) mode $k_2$ when it reaches the point where $U=U^*$, and is subsequently re-absorbed by the black hole. Note that the counter-propagating character of the mode with respect to a background flow is essential for this behaviour, just like a boomerang must be thrown against the wind direction (typically at 45$^\circ$) in order to come back.

We call the turning point where $U=U^*$ a ``red horizon'' in line with the term ``blue horizon'' introduced in~\cite{Rousseaux:2010md}: the red horizon is a turning point for red-shifted waves, and in the whole process at the red horizon the wave undergoes further redshifting. We could also reverse the time direction in Fig.~\ref{Fig:ray-plots} to study what happens for a mode trying to enter a white hole. In that case, the blocking line or turning point would be associated with a blue-shifting process, and we would recover a blue horizon in precisely the same sense as introduced in~\cite{Rousseaux:2010md}. A red horizon in the black-hole case thus corresponds (by time-inversion) to a blue horizon in the white-hole case. Note that the location of such a red/blue horizon is automatically frequency-dependent, as we discuss below, even though no higher-order terms in $k$ have been introduced in the dispersion relation.

\begin{figure*}[h] 
\includegraphics[width=11cm]{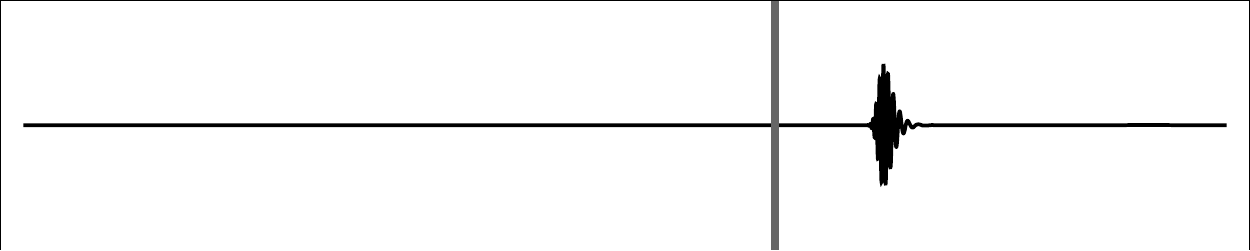} 

\vspace{1mm}

\includegraphics[width=11cm]{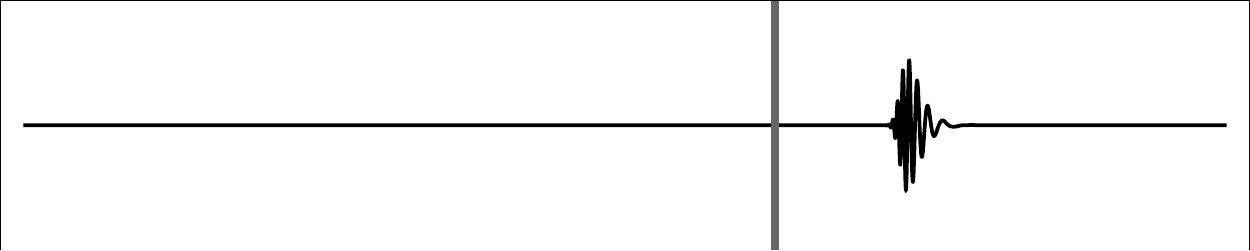}

\vspace{1mm}

\includegraphics[width=11cm]{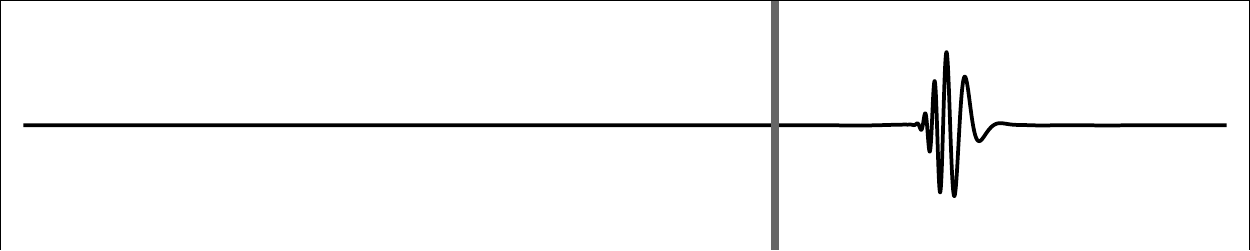}

\vspace{1mm}

\includegraphics[width=11cm]{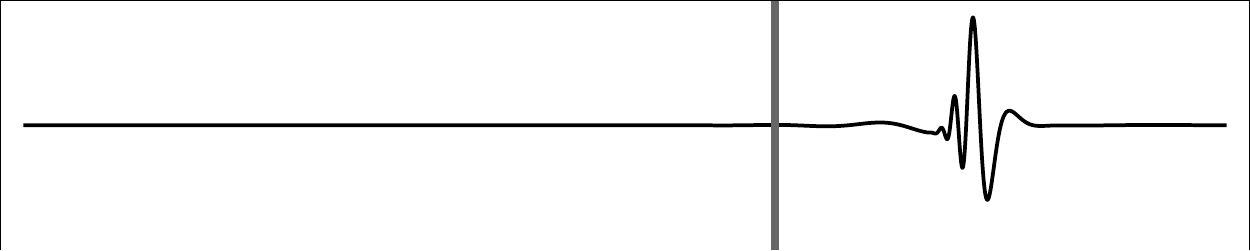}

\vspace{1mm}

\includegraphics[width=11cm]{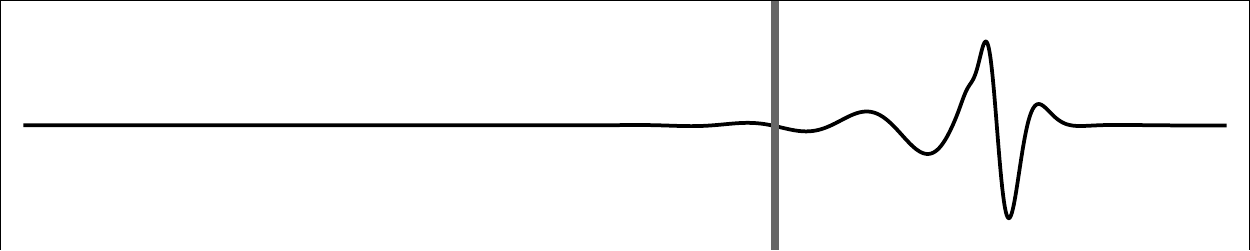}

\vspace{1mm}

\includegraphics[width=11cm]{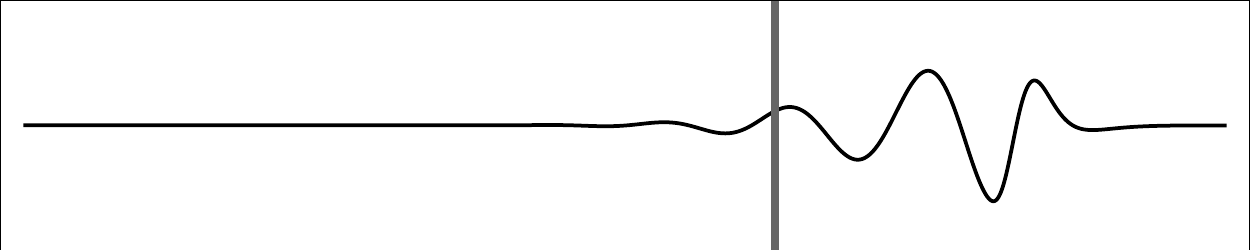}

\vspace{1mm}

\includegraphics[width=11cm]{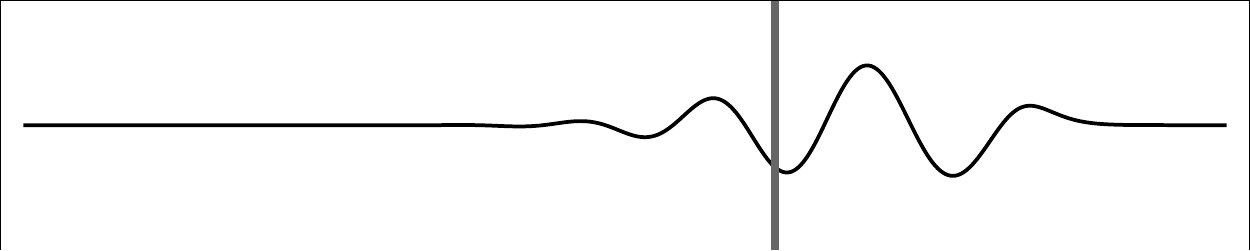}

\vspace{1mm}

\includegraphics[width=11cm]{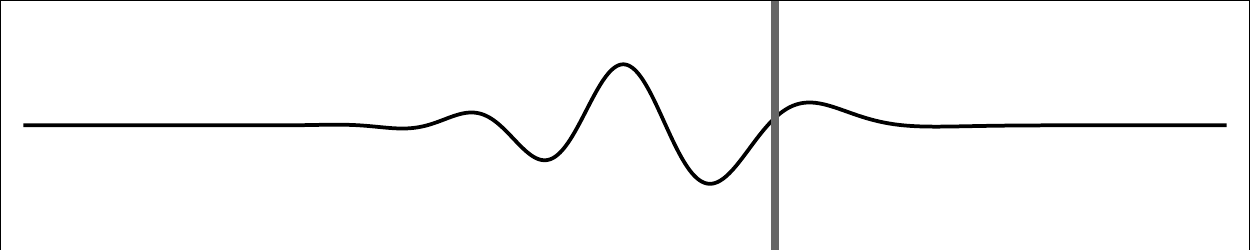}

\vspace{1mm}

\includegraphics[width=11cm]{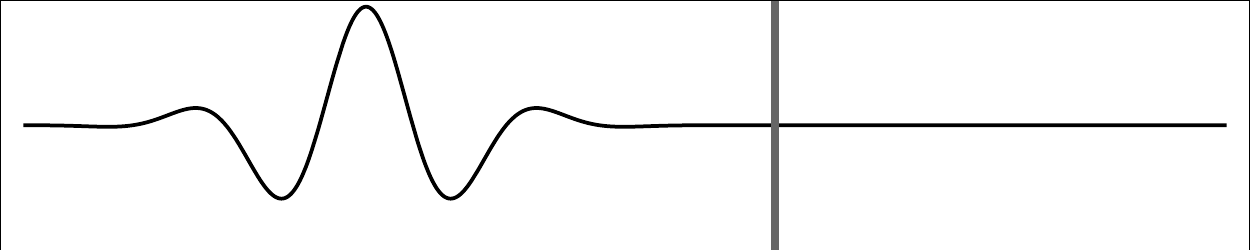}
\caption{Evolution of a Klein-Gordon wave packet with frequency spectrum contained in the range $0<\omega<m$. The grey vertical line is the horizon of a black hole (interior on the left, exterior on the right). The packet starts (top) close to the horizon, propagates away from the black hole, bounces and falls into the black hole interior, red shifting all the way.}
\label{Fig:num}
\end{figure*}

The bouncing behaviour of the green ray in the top plot in Fig.~\ref{Fig:ray-plots} ($0<\omega<m$), together with the continual red-shifting in time, is shown by a numerical solution for a Klein-Gordon wave packet centered on this ray, see Fig.~\ref{Fig:num}. The wave packet at various points in time is shown in  Fig.~\ref{Fig:num}, with time increasing from top to bottom. Initially the wave packet is outside the horizon (the grey vertical line) and moves to the right, away from the black hole. The packet reaches a maximum distance from the horizon before reversing direction and falling through the horizon into the black hole. The extreme red-shifting of the wave packet that accompanies its motion is evident. (The numerical simulation was performed backwards in time, beginning with the packet inside the black hole and solving for its history; equivalently, the simulation was performed forwards in time for the white hole obtained by reversing the flow $U$.)

\begin{figure} 
\includegraphics[width=.45\textwidth]{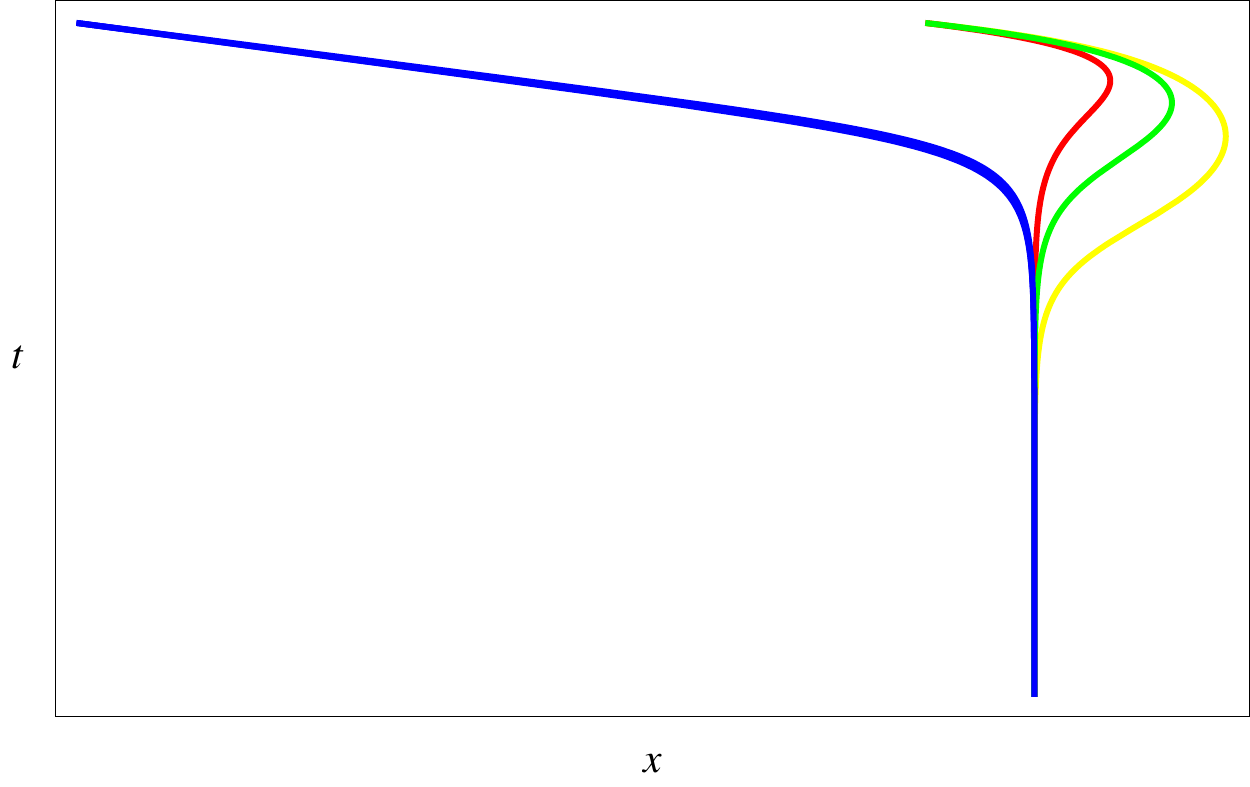}
\includegraphics[width=.45\textwidth]{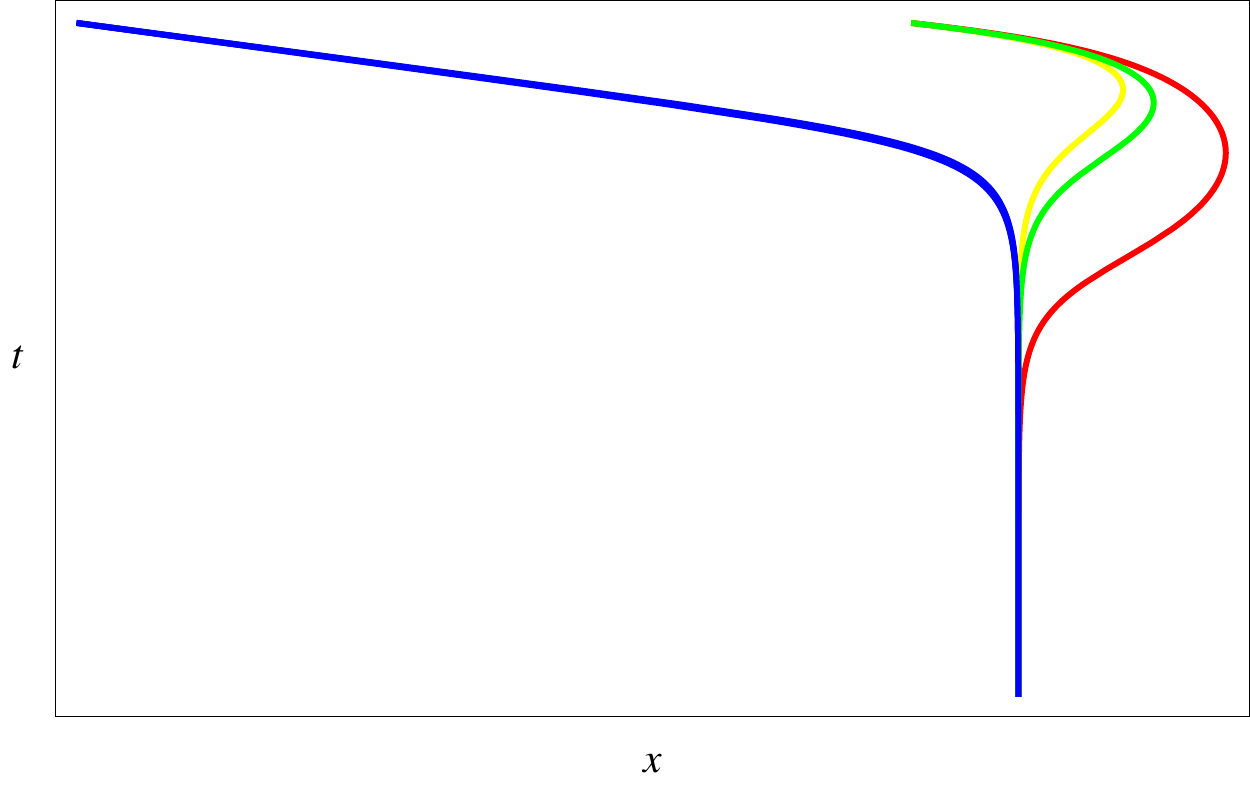}
\caption{\label{Fig:m-w-change} 
Mass- and frequency-dependence of the ``red horizon''. Mass (top): $m_\text{red}>m_\text{green}>m_\text{yellow}$. The heavier the mode, the less distance it can escape the black hole before being bounced back and reabsorbed. 
Frequency (bottom): $\omega_\text{red}>\omega_\text{green}>\omega_\text{yellow}$. The higher the frequency, the further towards  asymptotic Minkowski space the mode can escape. The green curve in both graphics corresponds to the values used in Fig.~\ref{Fig:ray-plots}.
}
\end{figure}

\subsection{Change of red horizon with changing mass or frequency}
The evolution of the red horizon with mass $m$ and frequency $\omega$ is shown in Fig.~\ref{Fig:m-w-change}. Note that the existence of the red horizon assumes $0<\omega<m$.

When the mass $m$ is increased, the position of the red horizon moves closer to the black-hole horizon. Conversely, if $m$ is decreased (keeping $m>\omega$), the position of the red horizon moves away from the black-hole horizon. This is easy to see from the value of the critical $U^*$ which identifies the red horizon, see eq.~(\ref{U*}): as $m$ increases, $|U^*|$ increases, and so lies closer to $|U|=c$, i.e. to the black-hole horizon. A massive mode with $m\to \infty$ will not be able to escape any distance from the horizon, since $|U^*|\to c$. In the lower limit $m \to \omega$, on the other hand, we obtain $U^*\to 0$, i.e. the mode can asymptotically reach infinity. 

If $\omega$ is increased (keeping $\omega<m$), the position of the red horizon moves further from the black-hole horizon. Conversely, if $\omega$ is decreased, the position of the red horizon moves closer to the black-hole horizon.
This is again easy to see from eq.~(\ref{U*}). In the limit $\omega \to 0$, the mode cannot escape any distance from the horizon ($|U^*|\to c$), whereas in the limit $\omega \to m$, we obtain $|U^*|\to 0$, so that the mode escapes to infinity.

\begin{figure} 
\includegraphics[width=9cm]{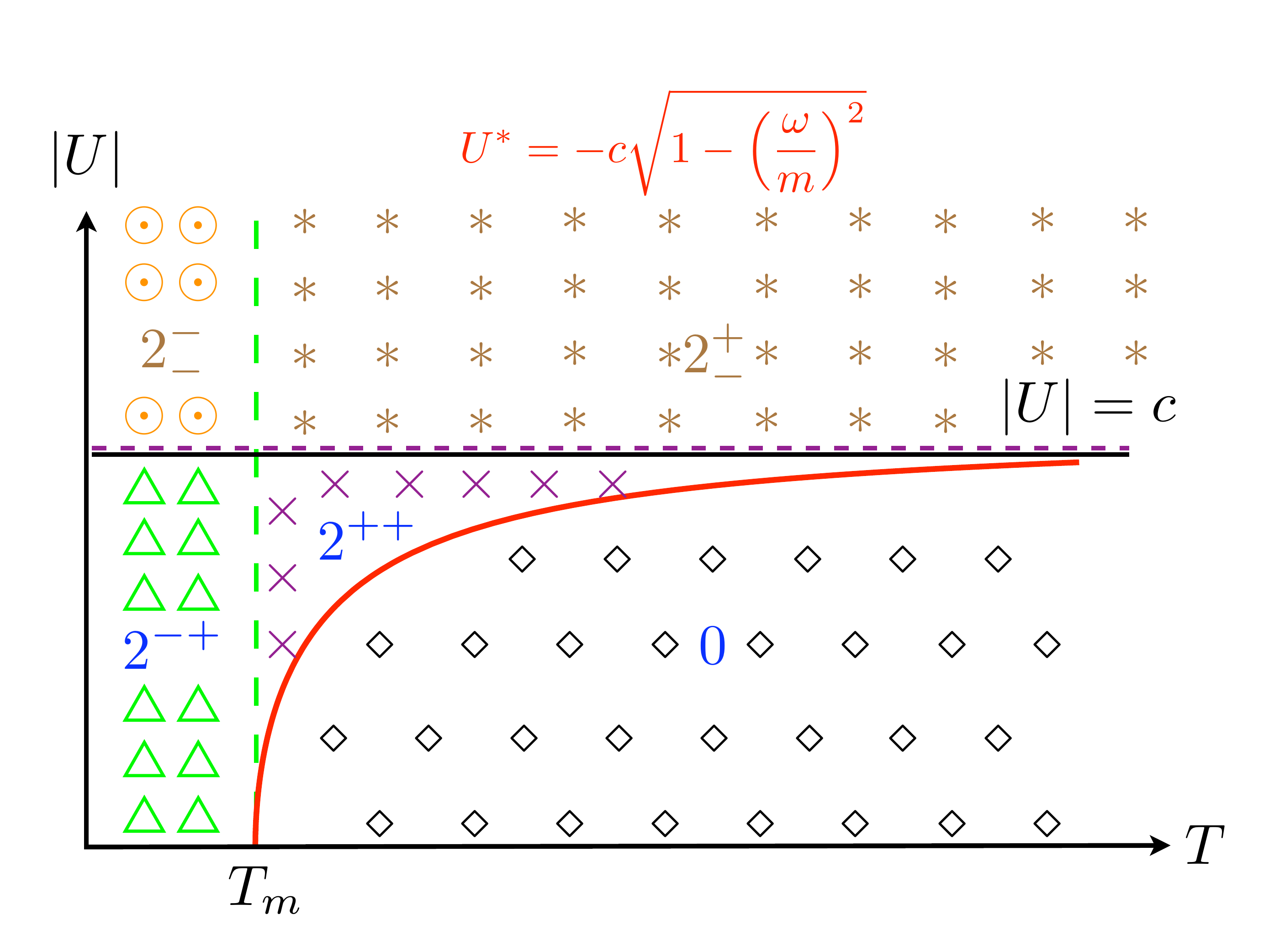}
\caption{\label{Fig:UvsT} 
Phase diagram of flow speed $|U|$ versus period $T=2\pi/\omega$. The purple horizontal dashed line is $|U|=c$ and corresponds to the black hole horizon. The red line shows the critical speed $|U^*|$, which defines the red horizon. The vertical dashed green line corresponds to $\omega=m$, i.e. $T_m=2\pi/m$. For $T<T_m$, everything still behaves qualitatively as in the massless case. For $T>T_m$, a second turning point of zero group velocity appears, the red horizon. Massive modes can only exist for $|U|\geq |U^*|$. The numbers in the phase diagram represent the number of roots ($k$'s) in each region and their signs; the upper position for a sign means the root has positive (co-moving) frequency, the lower position means the root has negative frequency.
}
\end{figure}

\section{Phase diagram}
Some of the preceding considerations can be summarised in a phase diagram of flow speed $|U|$ and period $T=2\pi/\omega$, see Fig.~\ref{Fig:UvsT}. The purple horizontal dashed line in Fig.~\ref{Fig:UvsT} is $|U|=c$, which represents the black-hole horizon. It also represents a ``negative horizon''~\cite{Rousseaux:2010md}: negative-frequency modes can exist only for $|U|>c$. This negative horizon is indicated by a horizontal black line in Fig.~\ref{Fig:UvsT}; although the negative horizon coincides with the black-hole horizon in the case discussed here, these two horizons can separate, for example in the case of surface waves on moving water~\cite{Rousseaux:2010md}. 

For massless modes, the purple (and black) line would strictly mark the only blocking line for modes: all counter-propagating massless modes have a positive group velocity $d\omega/dk>0$ in the lab frame as long as $|U|<c$ (they overcome the background flow), but a negative $d\omega/dk$ for $|U|>c$ (they are dragged along by the background flow). $|U|=c$ therefore marks the only blocking line ($d\omega/dk=0$) for modes in the massless case.

For massive modes, the picture becomes more complicated. The red line in Fig.~\ref{Fig:UvsT} shows the critical speed $|U|=|U^*|$ at the red horizon as a function of $T$. The vertical dashed green line corresponds to $\omega=m$, i.e. $T=T_m=2\pi/m$. As $T$ approaches $T_m$ from above the red horizon moves further away from the black-hole horizon $|U|=c$ and reaches spatial infinity at $T=T_m$, in line with the results of the last section. For $T<T_m$, the red horizon has disappeared and the behaviour is qualitatively the same as the massless case: counter-propagating modes have a positive lab-frame group velocity $d\omega/dk>0$ when $|U|<c$, but are dragged along by the background flow ($d\omega/dk<0$) when $|U|>c$. The only blocking line ($d\omega/dk=0$) for $T<T_m$ is therefore $|U|=c$. 
For $T>T_m$, however, a second blocking line $|U|=|U^*|$ is present, the ``red horizon'': for a given period $T$ (frequency $\omega$), massive modes can only exist for $|U|>|U^*|$, i.e. when the counterflow is sufficiently strong. For weaker background flows $|U|<|U^*|$, their existence is prohibited by the dispersion relation (\ref{disp-rel-squared}). As the mode moves away from the horizon ($|U|$ decreases below $c$), it is blocked ($d\omega/dk=0$) at the red horizon $|U|=|U^*|$, bounces back ($|U|$ increases again), crosses the black-hole horizon ($|U^*|=c$) and is absorbed by the black hole.

The numbers on the phase diagram represent the number of roots (values of $k$) and their signs; the positions of the signs show whether the root is positive (co-moving) frequency (upper position) or negative frequency (lower position). For example, a mode with $T>T_m$ has 0 roots for $|U|<|U^*|$, one (double) root $k>0$ for $U=U^*$ and two positive roots ($k_1, k_2>0$) for $|U^*|<|U|<c$. When $|U|\to c^-$, $k_1 \to +\infty$, and reappears as $k_1 \to -\infty$ when $|U|\to c^+$. For $|U|>c$, there is therefore one negative and one positive root. For $T<T_m$, there are 2 roots, one positive ($k_1$, counter-propagating) and one negative ($k_2$, co-propagating), both with positive (co-moving) frequency, as long as $|U|<c$. For $|U|>c$, the $k_1$ solution has converted into a negative-frequency and negative-$k$ root and therefore $k_1,k_2<0$. As before, the transition from $k_1>0$ to $k_1<0$ goes through $k_1\to \pm \infty$ for $|U|\to c^\mp$, i.e.: $k_1$ effectively disappears for $|U|=c$.

\section{Hawking radiation}\label{S:HR}
As was seen in the foregoing, the behaviour of all massive (counter-propagating) modes, including those with $\omega<m$, is identical very close to the black-hole horizon: they all experience the familiar infinite blue-shifting. This behaviour at the horizon is in fact a serious problem for the derivation of the Hawking effect for real black holes, as it invalidates the assumption of negligible back-reaction on the black hole. Nevertheless it is now well-established that the introduction of non-linear dispersion at high wave numbers does not destroy the Hawking mechanism (although see~\cite{Barcelo:2008qe,Barcelo:2010bq} for some possible complications with respect to the vacuum selection in the case of super-luminal dispersion), while it limits the blue-shifting at the horizon. This removal of infinite (co-moving) frequencies through dispersion is certainly present in all analogue systems and gives a robust basis for investigations of Hawking radiation. As all massive modes are not distinguished by their behaviour at the horizon, where the Hawking mechanism occurs, any fields described by the Klein-Gordon equation will exhibit the Hawking effect at all frequencies $\omega$, i.e. without a cutoff given by the mass. 

But the fact that a particular mode exhibits the Hawking effect does not imply that it contributes to the evaporation of the black hole. After being emitted from the black hole, particles with $\omega <m$ eventually turn around and fall though the horizon, so they do not contribute to the evaporation. But one could imagine capturing the $\omega <m$ particles before they fall back into the black hole, in which case they would contribute to the shrinking of the horizon. A detector $D$ placed sufficiently close to the black hole will detect a significant amount of massive particles emitted by the black hole that are not detected by a detector at infinity. The same detector $D$ would also register massive Hawking particles falling back into the black hole. 

For a Schwarzschild black hole, the position $x^*$ of the red horizon is obtained from $(U^*/c)^2=R_S/x^*$, with $R_S$ the Schwarzschild radius and $U^*$ given by eq.\eqref{U*}. This leads, for example, to $x^*\approx 3R_S$ for $\omega=0.8m$, and $x^*\approx 50R_S$ for $\omega=0.99m$. The spectral resolution of a hypothetical particle detector placed at a position $x_D$ will essentially be determined by the fly-by time of the emitted mode: the time between crossing the detector on its way out and its disappearance behind the black hole horizon, after having bounced on the red horizon. To obtain a sufficient resolution, one would need a supermassive black hole, and/or place the detector extremely close to the black hole horizon, where the mode is slowed down by the black hole's gravitational pull. In that region, all sorts of additional astrophysical effects would probably put strong limits on the detectability of the boomerang behaviour. Although these issues are not very different from the general problems of detecting (massless) Hawking radiation, it is clear that such considerations are of theoretical interest only in the astrophysical case. 

However, for an analogue system, there is no issue with the spectral resolution since the modes can perfectly well be detected inside the horizon. For an analogue system in which a field obeys the Klein-Gordon equation, the lack of a mass cutoff in the emission of Hawking radiation is very important, and we will dedicate the next section to it.

Explicit quantitative estimates for the detection rate $W$ of massive particles with $k_B T_H\ll \omega<m$ (with $T_H$ the usual Hawking temperature) can be obtained through the semiclassical tunneling formalism~\cite{Volovik:1999fc,Volovik:2009eb}. This gives the standard Hawking result for massless particles, modulated by a factor which depends on the particle mass $m$, the difference $m^2-\omega^2$, the distance $x_D$ to the detector, and the black-hole spacetime profile $U(x)$. While the general expression is rather involved, it strongly simplifies when $x_D\gg x_H$, with $x_H$ the radius of the horizon. In that case, one obtains the simple approximation 
\begin{equation}
W(\omega)\propto \exp\left[-\frac{\omega}{k_B T_H}\right]\exp\left[-\frac{2x_D}{c}\sqrt{m^2-\omega^2}\right]
\end{equation}
(in units such that $\hbar=1$, as before), with $T_H=|dU/dx|_{x=x_H}/2\pi$ the standard Hawking temperature. Compared to the standard case, the probability that a massive particle with $\omega<m$ arrives at a distance $x_D$ decreases exponentially with $x_D$, as expected.

\section{Analogue gravity}
\label{S:analogue-gravity}
The study of Hawking radiation in analogue-gravity systems opens up possibilities that are excluded in astrophysics. One can consider black- and white-hole horizons, which both produce Hawking radiation, and indeed combinations of horizons that communicate with each other (for example, the so-called black-hole laser~\cite{BHlaser}). One can also perform measurements on both sides of a horizon, so detection of both particles in a Hawking pair and measurement of their correlation becomes possible.

The study of astrophysical black holes may lead to the simplistic view that there is a lower frequency (energy) threshold in the production of massive particles by horizons. This in turn would lead to the view that an analogue system where a field described by the Klein-Gordon equation experiences a horizon is a poor choice for observing the Hawking effect. In reality the Hawking radiation produced by the horizon is in no way suppressed by the mass, and such an analogue system is in this regard just as promising a candidate as one in which the field obeys the massless wave equation. The $\omega<m$ particles would be just as accessible to detection as the
$\omega>m$ particles. In the case of a black-hole horizon the $\omega <m$ particles end up going into the ``black hole'', whereas the $\omega >m$ particles propagate away from the other side of the horizon, but all can be detected. For a white-hole horizon, what happens depends on how the dispersion relation is modified at high $k$ in the particular analogue system. All real systems will be dispersive at the high $k$ values experienced by modes near the horizon and instead of sticking at the white-hole horizon the rays will drift away from the horizon on one side or the other depending on whether the dispersion is super-luminal or sub-luminal at high $k$. Again, there is no problem in principle in detecting all the particles, for all $\omega$, in this white-hole case.

One may wonder whether the boomerang effect will not be mixed with the---possibly very complicated---dispersive high-$k$ behaviour near the horizon. Two elements should be borne in mind.
The first is that detailed studies based on rather general (mainly ``subluminal'', see e.g.~\cite{Unruh:1994je}) dispersion relations as well as for concrete systems with massless waves (BECs, surface waves, etc.) have shown that such potential ultraviolet problems do not spoil the Hawking effect and a thermal spectrum can be recovered under quite general conditions on the velocity profiles. For BECs, for example, this is demonstrated starting from the Bogoliubov--de Gennes equations~\cite{Macher:2009nz}, and the outcome can safely be expected to be independent of the masslessness or massiveness of the phonons. The second key point is that the boomerang effect is detectable as long as the background flow velocity $U$ is non-zero. In analogue systems, this region can be extended to a distance far beyond the immediate surroundings of the horizon, so that any complicated high-$k$ behaviour has been completely smoothened out.

It remains to identify analogue-gravity systems where these ideas could be tested experimentally. In most analogue systems that have been considered the waves obey the massless wave equation at low wave-numbers (with modifications at high wave-numbers). An exception is a recent study of acoustic waves in a rotating ion ring~\cite{Horstmann:2009yh,Horstmann:2010xd} where the discreteness of the ions leads to a dispersion relation at low $k$ that is not of the massless form, but these waves are not Klein-Gordon waves. We give three examples of systems where waves obey the Klein-Gordon equation and where the creation of horizons for the waves is possible.

Bose-Einstein condensates (BECs) are among the most discussed systems for analogue gravity. Theoretically, they provide the advantage of being well-studied and rather well-understood, while experimentally they are clean and quite flexible. Black hole configurations in a BEC were recently realized for the first time~\cite{steinhauer}. The phonons that are obtained in standard BECs are massless, but in~\cite{Visser:2005ss} a scheme was described for producing a ``massive" phonon in  a two-component BEC. In the presence of a  horizon created by a sub- to super-sonic flow these massive phonons should then behave as described above (with an appropriate modification at high $k$). 

The second example is Langmuir waves in a moving plasma~\cite{plasma}. These waves have a dispersion relation of the form $(\omega-Uk)^2=\omega_p^2 + \omega_p^2R_D^2k^2$, with $\omega_p$ the plasma frequency and $R_D$ the Debye radius. This is the Klein-Gordon dispersion relation with the cut-off frequency $\omega_p$ playing the role of mass. One obvious undesirable feature in this system is Landau damping~\cite{plasma} of the waves, but all analogue-gravity systems present their own challenges.

The third example is from barotropic waves~\cite{barotropic} in fluid mechanics. Barotropic waves are an assortment of waves in fluids with rotation, one type of which, known as inertia-gravity or Poincar\'{e} waves, has the Klein-Gordon dispersion relation~\cite{barotropic} $\omega^2=f^2+ghk^2$, where $f$ is the Coriolis parameter, $g$ the gravitational acceleration and $h$ is the fluid depth. As in the previous examples, a horizon can be created by a fluid flow.

The presence of a lower frequency cutoff in the flat (but not curved) space-time dispersion relation is the feature of Klein-Gordon waves that leads to the interesting behaviour we have discussed in this paper. This lower cutoff is not unique to Klein-Gordon waves and we note another possible analogue-gravity system with waves that exhibit a lower frequency cutoff, namely one based on spin waves. Spin waves~\cite{spinwaves} are propagating perturbations of the ordering in magnetic systems. These waves can have a variety of dispersion relations, depending on their particular type and propagation direction. The key points relevant to our present discussion are the following. First, several of these dispersion relations have the form $\omega=\pm(A_1 + A_2k^2)$, where $A_1,A_2>0$; there is therefore a lower frequency cutoff $A_1$. Second, a Doppler shift of spin waves can be induced by the application of an electric current~\cite{spindoppler0}, an effect that was demonstrated a few years ago~\cite{spindoppler}. In the presence of an electric current the dispersion relation becomes $\omega-Uk=\pm(A_1 + A_2k^2)$, with a doppler-shifted frequency $\omega-Uk$, where $U$ is proportional to the current density~\cite{spindoppler0}. One can therefore envisage creating horizons for spin waves using an electric current density that varies along the length of the sample due to a changing transverse area. Further details of this proposal will be given elsewhere. 

\section{Conclusions}
The Hawking effect for massive Klein-Gordon modes does not exhibit a lower frequency cutoff given by the mass. This does not contradict the obvious fact that black holes cannot emit massive particles with $\omega>m$ to asymptotic infinity. We have shown in detail that massive modes with $\omega<m$ display a boomerang behaviour: after being emitted from the horizon they eventually turn around and are dragged back into the black hole. This bouncing, or blocking, behaviour is a familiar feature in analogue gravity in the context of waves with sub-luminal or super-luminal dispersion (or a combination of both~\cite{Rousseaux:2010md}) at high wave numbers. In the Klein-Gordon case the interesting dispersive effects are at low wave numbers, but the same techniques of analysis reveal the behaviour. 

The lesson that the Hawking effect for Klein-Gordon modes is not suppressed compared to the massless case is an important one for analogue gravity. Hawking radiation of Klein-Gordon modes in an analogue system is just as amenable to experimental detection as radiation of massless modes. Analogue-gravity systems with Klein-Gordon waves are possible using two-component BECs~\cite{Visser:2005ss}, Langmuir waves in plasmas and barotropic waves in fluid mechanics.

\section*{Acknowledgements}
We thank Serena Fagnocchi, Carlos Barcel\'{o}, Luis Garay and 
Renaud Parentani for commenting on a draft of this manuscript. We especially 
thank Grisha Volovik for his suggestions w.r.t.\ semiclassical tunneling. Gil Jannes thanks INSMI, CNRS and the University of Nice for financial support. Thomas Philbin thanks the Royal Society of Edinburgh and the Scottish Government for financial support, and the University of Nice for a visiting appointment.


\end{document}